\documentstyle[aps,prb,epsf,twocolumn,floats]{revtex}
\begin{document}
\draft

\twocolumn[\hsize\textwidth\columnwidth\hsize\csname @twocolumnfalse\endcsname

\title{Out-of-Equilibrium Kondo Effect: Response to Pulsed Fields}
\bigskip
\author{Avraham Schiller$^1$ and Selman Hershfield$^2$}
\address{ $^1$ Racah Institute of Physics, The Hebrew University,
               Jerusalem 91904, Israel \\
          $^2$ Department of Physics and National High Magnetic
               Field Laboratory, University of Florida,\\
               PO Box 118440, Gainesville, FL 32611}
\date{\today}
\maketitle

\begin{abstract}
The current in response to a rectangular pulsed bias potential is
calculated exactly for a special point in the parameter space
of the nonequilibrium Kondo model. Our simple analytical solution
shows the all essential features predicted by the non-crossing
approximation, including a hierarchy of time scales for the rise,
saturation, and fall-off of the current; current oscillations with
a frequency of $eV/\hbar$; and the instantaneous reversal of the
fall-off current for certain pulse lengths. Rich interference patterns
are found for a nonzero magnetic field (either dc or pulsed), with
several underlying time scales. These features should be observable
in ultra small quantum dots.
\end{abstract}

\bigskip

]
\narrowtext

The recent observation of the Kondo effect in ultra small quantum
dots~\cite{QD1,QD2,QD3,QD4} has focused considerable attention on
the Kondo effect in mesoscopic systems. Acting as tunable Anderson
impurities, quantum-dot devices offer an outstanding opportunity
to test some of the basic notions of the Anderson impurity
model.~\cite{Anderson61} For example, the crossover from the
local-moment to the mixed-valent and empty-impurity
regimes,~\cite{QD1} the $\pi/2$ phase shift associated with
the Kondo effect,~\cite{Oreg2000} and the evolution of the
Kondo effect with dc~\cite{AA66,Kondo_DC,us_dc}
and ac~\cite{Kondo_AC,us_ac,QD_ac} bias potentials.
Most recently, an intriguing scenario for the response of a
quantum dot to a rectangular pulsed bias potential was put
forward by Plihal {\em et al.}~\cite{Plihal_et_al} Using a
time-dependent version of the non-crossing approximation,
these authors predicted a hierarchy of time scales for the
rise, saturation, and fall-off of the current through
the dot, featuring a rich interplay between the Kondo temperature,
the applied voltage bias, and the temperature. For larger
values of the pulsed bias $eV$, oscillations were predicted
in the time-dependent current, with a characteristic frequency
of $\omega = eV/\hbar$. These current oscillations were attributed
to the sharp excitation frequency between the two split
Kondo peaks induced by the $eV$ bias potential.~\cite{WM94}

In this paper, we exploit an exactly solvable nonequilibrium
Kondo model~\cite{us_dc,us_ac} to obtain a simple analytic
description of the Kondo-assisted current in response to
rectangular pulsed fields. Our solution captures all the
essential features reported by Plihal {\em et al.},~\cite{Plihal_et_al}
including the hierarchy of time scales for the rise, saturation,
and fall-off of the current; the $eV/\hbar$ oscillations
in the time-dependent current; and the instantaneous
reversal of the current -- for certain pulse lengths --
after the pulse has ended. Furthermore, we are able to
include also a nonzero magnetic field which itself can be
pulsed, revealing complex interference patterns between the
bias and magnetic field. Most importantly, since our solution
is analytic and exact, it provides a benchmark result for
the response of a quantum dot to pulsed fields.

We begin with a few general remarks on the solvable model
used. An extension of the Toulouse~\cite{Toulouse70} and
Emery-Kivelson~\cite{EK92} solutions of the equilibrium
Kondo problem to nonequilibrium,~\cite{Weiss} this model
is expected to correctly
describe the strong-coupling regime of the nonequilibrium
Kondo effect. Indeed, previous applications of the
model to dc~\cite{us_dc} and ac~\cite{us_ac} bias
potentials have shown all the qualitative features of
Kondo-assisted tunneling: A zero-bias anomaly that splits in an
applied magnetic field; Fermi-liquid characteristics in the
low-$T$ and low-$V$ differential conductance; Side peaks in
the differential conductance at $eV = \pm n\hbar \omega$,
for an ac drive of frequency $\omega$. At the same time,
since the solvable point involves strong couplings, this
model is incapable of describing weak-coupling features
such as the logarithmic temperature dependence of the
conductance at elevated $T$.

Introducing the actual physical system, it consists of two noninteracting
leads of spin-$1/2$ electrons, interacting via a spin-exchange
coupling with a spin-$1/2$ impurity moment $\vec{\tau}$ placed
in between the two leads. The impurity moment can represent either
an actual magnetic impurity or a quantum dot with a single
unpaired electron. Each lead is subject to a separate time-dependent
potential, such that a time-dependent voltage bias forms
across the junction. Since the
interaction with the impurity is local in space ($s$-wave
scattering), one can reduce the conduction-electron degrees of
freedom that couple
to the impurity to one-dimensional fields $\psi_{\alpha \sigma}(x)$,
where $\alpha = R, L$ labels the lead (right or left) and
$\sigma = \uparrow, \downarrow$ specifies the spin orientations.
In terms of these one-dimensional fields, coupling to the
impurity takes place via the local spin densities
$\vec{s}_{\alpha \beta} = \frac{1}{2}
\sum_{\sigma,\sigma'} \psi^{\dagger}_{\alpha \sigma}(0)
\vec{\sigma}_{\sigma, \sigma'} \psi_{\beta \sigma'}(0)$.

As discussed in detail in Refs.~\onlinecite{us_dc} and
\onlinecite{us_ac}, the exactly solvable model corresponds
to a particular choice of the spin-exchange couplings, described
by the Hamiltonian
\begin{eqnarray}
{\cal H} =&& {\cal H}_0
                 - \sum_{\alpha = L, R} eV_{\alpha}(t) \hat{N}_{\alpha}
                 - \mu_B g_i H(t)\tau^{z}\nonumber \\
          &&+ \sum_{\alpha, \beta = L, R}
                 J_{\perp}^{\alpha \beta} \left \{ s^x_{\alpha \beta} 
                 \tau^x + s^y_{\alpha \beta} \tau^y \right \} ,
\label{Full_H}
\end{eqnarray}
\begin{eqnarray}
{\cal H}_0  =  i\hbar v_{F}\sum_{\alpha = L, R} &&
               \sum_{\sigma = \uparrow,\downarrow}
               \int_{-\infty}^{\infty}\!\psi_{\alpha \sigma}^{\dagger}(x)
               \frac{\partial}{\partial x} \psi_{\alpha \sigma}(x) dx
\nonumber \\
+ && J_z \left ( s^z_{LL} + s^z_{RR} \right) \tau^z .
\label{unperturbed}
\end{eqnarray}
Here the inter-lead longitudinal couplings have been set equal to zero,
while the intra-lead longitudinal couplings take the value
$J_z = 2\pi\hbar v_F$. The three transverse couplings $J_{\perp}^{LL},
J_{\perp}^{RR}$, and $J_{\perp}^{LR} = J_{\perp}^{RL}$ are arbitrary.
$\hat{N}_{\alpha}$ and $V_{\alpha}(t)$ in the second term of
Eq.~(\ref{Full_H}) are the number operator and time-dependent
potential on lead $\alpha$, respectively, while $H(t)$ is a local
magnetic field acting on the impurity spin.

The basic approach we take is identical to that of
Ref.~\onlinecite{us_ac}, where the same system was
solved for ac drives. The first step is to convert the
interacting nonequilibrium problem to a noninteracting one.
To this end, we first decompose the Hamiltonian into an
unperturbed part and a perturbation, where the unperturbed
part consists of ${\cal H}_0$ and the time-dependent voltage
$V_{\alpha}(t)$. The initial density matrix --- characterizing
the distribution of the unperturbed system before tunneling
has been switched on --- is equal to
$\rho_0 = e^{-\beta {\cal H}_0}/{\rm Trace}
\left \{ e^{-\beta {\cal H}_0} \right \}$.

The Hamiltonian, its unperturbed part, and the initial density
matrix are then converted to quadratic forms via the following
sequence of steps:~\cite{us_dc,EK92} (i) Bosonizing the fermion fields;
(ii) Converting to new boson fields corresponding to collective charge
($c$), spin ($s$), flavor ($f$), and spin-flavor ($sf$) excitations;
(iii) Performing a canonical transformation; (iv) Refermionizing
the new boson fields by introducing four new fermion fields,
$\psi_{\mu}(x)$ with $\mu = c, s, f, sf$; and (v) Representing
the impurity spin (which has been mixed by the canonical
transformation with the conduction-electron spin degrees of
freedom) in terms of two Majorana fermions:
$\hat{a} = -\sqrt{2}\tau_y$ and $\hat{b} = -\sqrt{2}\tau_x$.
For $J_z = 2\pi \hbar v_F$, one thus arrives at the following
representation of the Hamiltonian:
\begin{eqnarray}
{\cal H}' &=& \sum_{\mu = c, s, f, sf} \sum_k \epsilon_k
         \psi^{\dagger}_{\mu,k}\psi_{\mu,k} - eV_{c}(t) \hat{N}_c
          + eV(t) \hat{N}_f
\nonumber\\
         && + i \mu_B g_i H \hat{b}\hat{a}
         + i \frac{J^{+}}{2\sqrt{\pi a L}}
         \sum_k ( \psi^{\dagger}_{sf,k} + \psi_{sf,k} ) \hat{b}
\nonumber\\
         && + \frac{J_{\perp}^{LR}}{2\sqrt{\pi a L}}
         \sum_k ( \psi^{\dagger}_{f,k} - \psi_{f,k} ) \hat{a}
\nonumber\\
         && + \frac{J^{-}}{2\sqrt{\pi a L}}
         \sum_k ( \psi^{\dagger}_{sf,k} - \psi_{sf,k} ) \hat{a} .
\label{h'}
\end{eqnarray}
Here $J^{\pm}$ are equal to $(J^{LL}_{\perp} \pm J^{RR}_{\perp})/2$;
the energies $\epsilon_k$ are equal to $\hbar v_F k$; $\hat{N}_c$
and $\hat{N}_f$ are the number operators for the charge and flavor
fermions, respectively; $V_{c}(t)$ and $V(t)$ are equal
to $V_L(t) + V_R(t)$ and $V_R(t) - V_L(t)$, respectively; $a$ is
an ultraviolet momentum cutoff, corresponding to a lattice spacing;
and $L$ is the effective size of the system (i.e., $k$ is discretized
according to $k = 2\pi n/L$).
Within this mapping, the Hamiltonian term ${\cal H}_0$ is reduced
to the free kinetic-energy part of Eq.~(\ref{h'}), ${\cal H}'_{kin}$,
and so $\rho_0$ is transformed to $e^{-\beta{\cal H}'_{kin}}/{\rm Trace}
\{ e^{-\beta{\cal H}'_{kin}}\}$.

\begin{figure}
\centerline{
\vbox{\epsfxsize=80mm \epsfbox {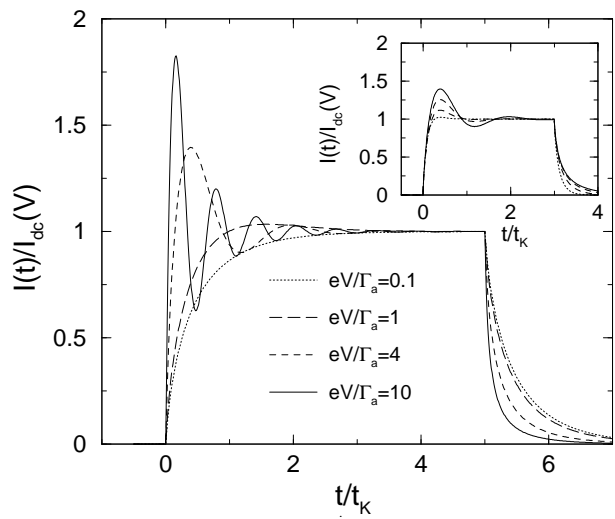}}
}
\caption{The time-dependent current $I(t)$ in response to an $eV$
bias potential that is switched on at time $t = 0$, and then
switched off again at time $t = \tau$. Here $\tau = 5 t_K$ with
$t_K = \hbar/\Gamma_a$, $k_B T$ is fixed at $0.1 \Gamma_a$, and
$I_{dc}(V)$ is the steady-state current for a dc voltage bias of
$eV$ [see Eq.~(\ref{I_dc})]. As $eV$ is increased, oscillations develop
in $I(t)$, leading to a significant reduction in the rise time.
The saturation time, however, is unaffected by the bias at the
solvable point. Inset:
Temperature dependence of $I(t)$ for fixed $eV/\Gamma_a = 4$
and $\tau/t_K = 3$. Full, long-dashed, dashed, and dotted lines
correspond to $k_BT/\Gamma_a = 0.1, 0.5, 1$ and $2$, respectively.
With increasing $T$, the saturation time is reduced and the
current oscillations are suppressed.}
\label{fig:fig1}
\end{figure}

Having converted the problem to a noninteracting form, one can sum
exactly all orders in the perturbation theory to obtain the current.
The solution features two basic energy scales,~\cite{us_dc,us_ac}
$\Gamma _a = [(J_{\perp}^{LR})^2 + (J^{-})^2] / 4 \pi a\hbar v_F$
and $\Gamma _b = (J^{+})^2/ 4\pi a\hbar v_F$, which determine
the width of the different Majorana spectral functions, and thus
play the role of Kondo temperatures.~\cite{EK92} The conventional
one-channel Kondo effect is best described by the case where
$\Gamma_a = \Gamma_b$, for which only a single Kondo scale
emerges.~\cite{us_dc}

Following Plihal {\em et al.},~\cite{Plihal_et_al} we begin
with the case of a pulsed bias potential and a zero magnetic
field. Explicitly, we assume that a bias potential of $eV$
is switched on at time $t = 0$, and then switched off again
at time $t = \tau$. Denoting 
$(J_{\perp}^{LR})^2 / 4 \pi a\hbar v_F$ by $\Gamma_1$, the resulting
time-dependent current $I(t)$ flowing from right to left is
conveniently expressed via the auxiliary function
\begin{eqnarray}
{\cal J}(t; V_1, V_2) =&&
         \frac{e\Gamma_1}{\pi \hbar} \sum_{n = 0}^{\infty}
         e^{-[(2n+1)\pi k_B T + \Gamma_a]t/\hbar} \times 
\label{J(t,v,v)} \\
&& \left(
         \frac{1}{n + \frac{1}{2} + \frac{\Gamma_a + ieV_1}{2\pi k_B T}} -
         \frac{1}{n + \frac{1}{2} + \frac{\Gamma_a + ieV_2}{2\pi k_B T}}
   \right) .
\nonumber
\end{eqnarray}
For $t \leq 0$, $I(t)$ is obviously zero. For $ 0 \leq t \leq \tau$
it equals
\begin{equation}
I_1(t) = I_{dc}(V) -
             {\rm Im} \left \{ e^{-ieV t/\hbar}
             {\cal J}(t; 0, V) \right \} ,
\label{I_1(t)}
\end{equation}
while for $\tau \leq t$ it is given by
\begin{equation}
I_2(t) = {\rm Im} \left \{ {\cal J}(t\!-\!\tau; 0, V) \right \}
         - {\rm Im} \left \{ e^{-ieV\tau/\hbar}
                 {\cal J}(t; 0, V) \right \} .
\label{I_2(t)}
\end{equation}
Here $I_{dc}(V)$ is the steady-state current for an $eV$
dc voltage bias:~\cite{us_dc}
\begin{equation}
I_{dc}(V) = \frac{e\Gamma_1}{\pi \hbar} {\rm Im} \left \{ \psi
              \left (\frac{1}{2} + \frac{\Gamma_a + ieV}{2\pi k_B T}
              \right ) \right \} .
\label{I_dc}
\end{equation}

The relevant time scales for the evolution of the current are easily
read off from Eqs.~(\ref{I_1(t)}) and (\ref{I_2(t)}). As proposed
by Plihal {\em et al.},~\cite{Plihal_et_al} for $0 \leq t \leq \tau$
the current has two components: a constant term equal to the new
steady-state current, and a transient term that oscillates with
frequency $\omega = eV/\hbar$. The latter component involves two
natural times scales: the period of oscillations $t_v = 2\pi \hbar/e V$,
and the basic decay time $t_{d} = \hbar/(\Gamma_a + \pi k_B T)$, which
enters via ${\cal J}(t, 0, V)$. For $t_d \ll t_v$, the current
saturates before any oscillations develop. Hence the rise and saturation
times are both of the order of $t_d$. On the other hand, for $t_v < t_d$
the current oscillates before saturating at $I_{dc}(V)$, resulting
in a much shorter rise time. Specifically, the rise time is identified
with the first instance at which the transient term vanishes.
Writing ${\cal J}(t; 0, V)$ of Eq.~(\ref{J(t,v,v)}) as
$|{\cal J}(t; 0, V)| e^{i\varphi(t)}$, it is simple to show
that $0 < \varphi(t) < \frac{\pi}{2}$. As the transient
current vanishes each time $eVt/\hbar - \varphi(t)$ equals an
integer multiple of $\pi$, this first happens at some time shorter
than $t_v/4$. Thus, while the saturation time still follows
$t_d$, the rise time is bounded by $t_v/4 \ll t_d$. This
transition in rise time from approximately $t_d$ to less than
$t_v/4 \ll t_d$ is clearly seen in Fig.~\ref{fig:fig1}, where the
time-dependent current is plotted for increasing values of $eV$.

Contrary to the current oscillations that may develop for
$0 \leq t \leq \tau$, the fall off for $t > \tau$ is described in
Eq.~(\ref{I_2(t)}) by the difference of two non-oscillatory terms,
each of which decays to zero. While the tail of the fall off is
always characterized by the fall-off time $t_d$, the initial
stages of the drop do not
typically follow a single time scale. This is best seen for long
pulse durations, \mbox{$\tau \gg t_d$}, when only the first term
survives in Eq.~(\ref{I_2(t)}). To this end, consider
${\rm Im} \{ {\cal J}(t-\tau; 0, V) \}$ with $t = \tau$ and
$t_v < t_d$. Using the series representation of
Eq.~(\ref{J(t,v,v)}), an increasing number of terms contribute
to ${\rm Im} \{ {\cal J}(0; 0, V) \}$ as $eV$ is increased,
typically up to $n \sim eV/2\pi k_B T$.
For $t > \tau$, each of these terms decays with a separate relaxation
time, $t_d^{(n)} = \hbar/[(2n + 1)\pi k_B T + \Gamma_a]$, such
that the total signal contains a range of time scales from $\sim t_v$
up to $t_d$. While all time scales participate in the early stages
of the fall off, the tail is governed by the longest time scale,
$t_d$.

\begin{figure}
\centerline{
\vbox{\epsfxsize=75mm \epsfbox {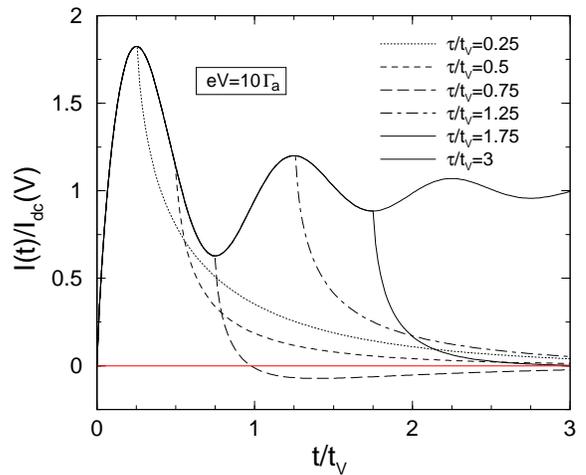}}
}
\caption{The current $I(t)$ in response to a pulsed bias potential of
magnitude $eV=10\Gamma_a$, for different pulse durations $\tau$. Here
$k_B T = 0.1\Gamma_a$ and $t_v = 2\pi\hbar/eV$. $I_{dc}(V)$ is the
steady-state current for an $eV$ dc voltage bias. The asymptotic
fall-off current oscillates with $\sin(2\pi \tau/t_v)$,
resulting in the reversal of the current for $\tau/t_v = 3/4$.}
\label{fig:fig2}
\end{figure}

For shorter pulse durations, $\tau \sim t_v < t_d$, the fall-off current
is actually quite sensitive to the precise value of $\tau$. This is
demonstrated in Fig.~\ref{fig:fig2}, where several pulse durations
are plotted for fixed $eV/\Gamma_a = 10$. Although
$eV/\Gamma_a = 10$ may appear too large a bias for the application
of the solvable point, the qualitative agreement with the numerical
calculations of Plihal {\em et al.} (see
Ref.~\onlinecite{Plihal_et_al}, Fig.~3) is strikingly good. In fact,
we are even able to reproduce the reversal of the fall-off current
for certain pulse lengths. To understand the origin of this
somewhat surprising effect, we go back to Eq.~(\ref{I_2(t)}) with a
sufficiently large $t - \tau$, such that each ${\cal J}$ function
can be roughly approximated by the $n = 0$ term in
Eq.~(\ref{J(t,v,v)}) (we assume $T > 0$). For a large voltage
bias such that $eV \gg \Gamma_a$ and $eV \gg \pi k_B T$ (as is
the case in Fig.~\ref{fig:fig2}), the fall-off current of
Eq.~(\ref{I_2(t)}) is asymptotically given by
\begin{eqnarray}
I_2(t) \sim \frac{2 k_B T \Gamma_1}{\hbar V} e^{-(t - \tau)/t_d} &&
       \left [ 1 - e^{-\tau/t_d} \cos \left( 2\pi \frac{\tau}{t_v}
       \right) + \right.
\nonumber\\
&& \left. \frac{2\pi t_d}{t_v} e^{-\tau/t_d}
       \sin \left (2\pi \frac{\tau}{t_v} \right) \right] .
\label{I_asymptotic}
\end{eqnarray}
Thus, since $t_d \gg t_v$ and $\tau \sim t_v$ under the assumptions
above, the asymptotic fall-off current is dominated by the
last term in Eq.~(\ref{I_asymptotic}). Consequently, the sign of
$I_2(t)$ oscillates with $\sin (2\pi \tau/t_v)$, precisely
as seen in Fig.~\ref{fig:fig2}.

Although our solution clearly captures the essential findings of
the non-crossing approximation, there are two major omissions to be
noted. First, working with a Kondo impurity rather than an Anderson
impurity, our model lacks the short charge-fluctuation time scale
($t_0 = \hbar/\Gamma_{\rm dot}$ in the notations of
Ref.~\onlinecite{Plihal_et_al}), which governs the very early
response of a quantum dot to the abrupt change in the applied
voltage bias. Second, for $0 < t \leq \tau$, our solution does not
show the expected decrease in saturation time for large voltage
bias.~\cite{Plihal_et_al} This decrease, which stems from the
dissipative lifetime induced by the bias potential,~\cite{WM94} is not
captured by the solvable point. On the other hand, our solution
gives a very transparent picture for the role of a temperature.
As the temperature is increased, $t_d$ crosses over from
$\hbar/\Gamma_a$ to $\hbar/\pi k_B T$, which remains the only
relevant time scale for $k_B T \gg eV$. Thus, all response times
for the current are determined by the temperature in this limit.

\begin{figure}
\centerline{
\vbox{\epsfxsize=70mm \epsfbox {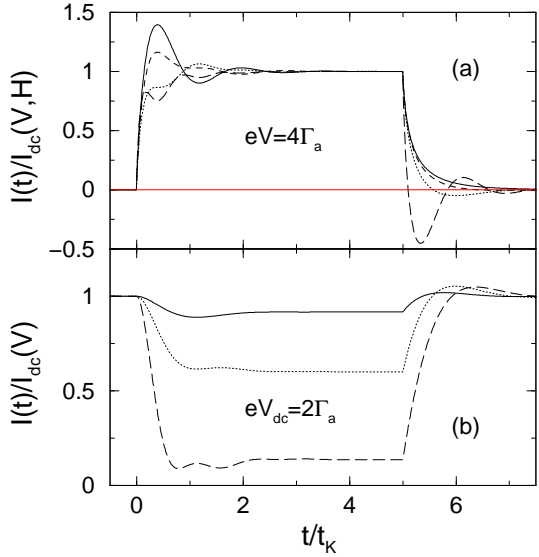}}
}
\caption{The time-dependent current for (a) a dc magnetic field $H$
and a rectangular pulsed bias potential of magnitude $eV=4\Gamma_a$;
and (b) a dc bias potential $eV=2\Gamma_a$ and a rectangular pulsed
magnetic field of magnitude $H$. Upper panel: Solid, dashed, dotted,
and long-dashed lines correspond to $\mu_B g_i H/\Gamma_a = 0, 1, 2$,
and $4$, respectively. Lower panel: Solid, dotted, and long-dashed
lines correspond to $\mu_B g_i H/\Gamma_a = 1, 2$, and $4$, respectively.
In both cases, the pulse duration is $\tau = 5t_k$
($t_k = \hbar/\Gamma_a$), and the temperature is equal to
$k_B T/\Gamma_a = 0.1$. $I_{dc}(V, H)$ marks the steady-state
current for a dc voltage bias of $eV$ and a dc magnetic field $H$.}
\label{fig:fig3}
\end{figure}

An important advantage of the solvable point with $\Gamma_a =
\Gamma_b$ is the ability to incorporate an arbitrary time-dependent
magnetic field. Specifically, we find that the current $I(t)$
for a general voltage bias $eV(t)$ and an arbitrary field $H(t)$
is equal to $\frac{1}{2} \left [ I_{+}(t) + I_{-}(t) \right ]$,
where $I_{\pm}(t)$ is the current for a zero magnetic field and
an effective bias potential $eV_{\pm}(t) = eV(t) \pm \mu_B g_j H(t)$.
The latter bias has a simple physical interpretation.
When an electron tunnels by flipping the impurity spin, the
effective potential barrier it sees is equal to $eV_{\pm}(t)$,
depending on the initial impurity state, and the direction
of tunneling.

Using the above decomposition of $I(t)$ into $I_{\pm}(t)$, we
have computed the current for the two opposite cases of (i) a
dc magnetic field $H$ and a rectangular pulsed bias potential
of magnitude $eV$, and (ii) a dc bias potential of $eV$ and
a rectangular pulsed magnetic field of magnitude $H$. In both
cases, the pulse duration was taken to be $\tau$, with the
corresponding field being equal to zero for $t < 0$ and
$t > \tau$. The calculation of $I_{\pm}(t)$ in each of these cases
requires the generalization of Eqs.~(\ref{I_1(t)})--(\ref{I_2(t)})
to a pulsed bias potential of the form $V(t) = V_2$ for
$0 \leq t \leq \tau$ and $V(t) = V_1 \neq 0$ otherwise,
which gives
\begin{eqnarray}
I_1(t) =&& I_{dc}(V_2) -
             {\rm Im} \left \{ e^{-ieV_2 t/\hbar}
             {\cal J}(t; V_1, V_2) \right \} , \\
I_2(t) =&& I_{dc}(V_1) + {\rm Im} \left \{ e^{-ieV_1 (t - \tau) /\hbar}
             {\cal J}(t\!-\!\tau; V_1, V_2)\right \}
\nonumber \\
&&- {\rm Im} \left \{ e^{-ieV_2 \tau/\hbar} e^{-ieV_1 (t - \tau) /\hbar}
             {\cal J}(t; V_1, V_2) \right \} .
\end{eqnarray}
For $t \leq 0$, the corresponding current is obviously $I_{dc}(V_1)$.

The resulting $I(t)$ curves are depicted in Fig.~\ref{fig:fig3}.
As is clearly seen, the combination of a bias and a magnetic field
produces nontrivial interference patterns with several underlying
time scales. These include $t_d$,
$t_{\pm} = 2\pi \hbar/|eV \pm \mu_B g_i H|$, and either
$t_h = 2 \pi \hbar/\mu_B g_i H$ or $t_v$, depending on whether
one is dealing with a pulsed bias potential or a pulsed magnetic
field. In particular, for a pulsed bias potential and moderately
large dc magnetic fields, there are current oscillations in the
fall-off current, with a characteristic frequency of
$\omega = \mu_B g_i H/\hbar$. Such effects should be observable
in ultra small quantum dots.

\end{document}